\documentclass[sigconf,screen]{acmart}

\acmConference[BDA]{Gestion de Données – Principes, Technologies et Applications}{20--23 October 2025}{Toulouse, France}
\acmCodeLink{https://github.com/Neplex/ArchiTXT}

\usepackage[T1]{fontenc}

\usepackage{hyperref}
\usepackage{graphicx}
\usepackage[linesnumbered,ruled]{algorithm2e}
\usepackage{forest}
\usepackage{amsmath}
\usepackage{amsthm}
\usepackage{amsfonts}
\usepackage{mathtools}
\usepackage{balance}
\usepackage[inline]{enumitem}
\usepackage{subcaption}
\usepackage{adjustbox}
\usepackage{pdflscape}
\usepackage{csquotes}
\usepackage{orcidlink}
\usepackage{booktabs}

\usepackage{pgfplots}
\usepackage{pgfplotstable}
\usepackage{wrapfig}
\usepackage{array}
\usepackage{todonotes}

\usepgfplotslibrary{units}
\usepgfplotslibrary{colorbrewer}
\usepgfplotslibrary{groupplots}
\usetikzlibrary{patterns}

\pgfplotsset{
    compat=newest,
    cycle list/Set1,
    scaled x ticks=false,
    scaled y ticks=false
}

\tikzset{
  labeled node/.style={
    rectangle split, rectangle split parts=2,
    rectangle split draw splits=false,
    rounded corners,
    very thick,
    draw=blue!50,
    rectangle split part fill={blue!50, white},
    align=left,
    font=\sffamily\bfseries\boldmath\small
  },
  labeled edge/.style={
    rectangle split, rectangle split parts=1,
    rectangle split draw splits=false,
    rounded corners,
    very thick,
    draw=orange!60,
    rectangle split part fill={orange!60, white},
    align=left,
    font=\sffamily\bfseries\boldmath\small
  },
  labeled property edge/.style={
    rectangle split, rectangle split parts=2,
    rectangle split draw splits=false,
    rounded corners,
    very thick,
    draw=orange!60,
    rectangle split part fill={orange!60, white},
    align=left,
    font=\sffamily\bfseries\boldmath\small
  },
  every two node part/.style={font=\sffamily\small},
}

\begin{document}

\newif\ifanonimized
\anonimizedfalse 

\newcommand{\selfCite}[1]{%
  \ifanonimized
    \cite{AN}%
  \else
    \cite{#1}%
  \fi
}

\newcommand{\githubCite}[0]{%
  \ifanonimized
    \cite{AN-GITHUB}%
  \else
    \cite{chabinarchitxt2024}%
  \fi
}

\ifanonimized
  \def\architxt{$\langle \text{anonymous} \rangle$\xspace}
\else
  \def\architxt{ArchiTXT\xspace}
\fi

\def\ie{\textit{i.e.}}
\def\eg{\textit{e.g.}}
\def\wrt{with respect to}
\def\redud{\rho}
\def\supp{\textsf{supp}}
\def\conf{\textsf{conf}}
\def\depScore{\delta}
\def\dup{\textsf{duplicates}}
\def\cov{\textsf{cs}} 
\def\comp{\textsf{cc}} 
\def\AMI{\textsf{AMI}}
\def\nbR{\textsf{\#R}}
\def\gO{\textsf{grOverlap}}

\def\metaG{\ensuremath{\mathcal{G}}\xspace}

\newcommand{\metamodelpart}[2]{\ensuremath{\textsf{#1}_{\text{#2}}\xspace}}
\newcommand{\metaProperty}[1]{\metamodelpart{Prop}{#1}}
\newcommand{\metaGroup}[1]{\metamodelpart{Grp}{#1}}
\newcommand{\metaRelation}[1]{\metamodelpart{Rel}{#1}}
\newcommand{\metaCollection}[1]{\metamodelpart{Coll}{#1}}

\theoremstyle{acmplain}
\newtheorem{axiom}{Axiom}

\theoremstyle{acmdefinition}
\newtheorem{remark}[theorem]{Remark}

\newcommand{\comm}[3]{\todo[inline,color={#2},caption={}]{\textbf{#1}: #3}}
\newcommand{\mhf}[1]{\comm{MHF}{orange}{#1}}
\newcommand{\jac}[1]{\comm{Jac}{yellow}{#1}}
\newcommand{\nh}[1]{\comm{Nicolas}{yellow}{#1}}

\newcommand*{\vertbar}{\rule[-1ex]{0.5pt}{2.5ex}}
\newcommand*{\horzbar}{\rule[.5ex]{2.5ex}{0.5pt}}

\newcommand*{\drawCircle}[1]{%
    \raisebox{-.5ex}{\tikz{\draw[fill=#1,line width=.5pt] circle[radius=1ex];}}%
}%

\title{Bridging Textual Data and Conceptual Models}
\subtitle{A Model-Agnostic Structuring Approach}

\ifanonimized
  \author{Anonymized Authors}
  \renewcommand{\shortauthors}{Anonymized}
\else
  \author{Jacques Chabin}
  \orcid{0000-0003-1460-9979}
  \affiliation{%
    \institution{Universit\'e d'Orl\'eans, INSA CVL, LIFO, UR 4022}
    \city{Orléans}
    \country{France}
  }
  \email{jacques.chabin@univ-orleans.fr}
  \author{Mirian Halfeld-Ferrari}
  \orcid{0000-0003-2601-3224}
  \affiliation{%
    \institution{Universit\'e d'Orl\'eans, INSA CVL, LIFO, UR 4022}
    \city{Orléans}
    \country{France}
  }
  \email{mirian@univ-orleans.fr}
  \author{Nicolas Hiot}
  \orcid{0000-0003-4318-4906}
  \affiliation{%
    \institution{Universit\'e d'Orl\'eans, INSA CVL, LIFO, UR 4022}
    \city{Orléans}
    \country{France}
  }
  \email{nicolas.hiot@univ-orleans.fr}
  \renewcommand{\shortauthors}{J. Chabin et al.}
\fi
\begin{abstract}
We introduce an automated method for structuring textual data into a model-agnostic schema, enabling alignment with any database model. It generates both a schema and its instance. Initially, textual data is represented as semantically enriched syntax trees, which are then refined through iterative tree rewriting and grammar extraction, guided by the attribute grammar meta-model \metaG. The applicability of this approach is demonstrated using clinical medical cases as a proof of concept.

\end{abstract}
\maketitle
\section{Introduction}
\label{sec:Intro}

The rapid growth of unstructured data, particularly textual data, offers valuable insights but presents challenges for analysis due to its complexity. 
In contrast, structured data, with its defined schema and relationships, allows for efficient analysis and easier reuse. 
When it comes to considering how to organize information that originally comes from a text, a critical consideration is the choice of an appropriate data model. Currently, the primary options typically range between relational da\-ta\-bases, offering consistency and NoSQL models providing flexibility. 
Complicating matters further, certain applications require interfacing with multiple data models simultaneously. 
A \textit{model-agnostic} solution enables seamless integration and interoperability. It provides a higher level of abstraction.

%
%

This paper presents \architxt, a tool for automatically structuring textual data into a model-agnostic database format. Our challenge is twofold. On one side, the \textit{automatic structuring} of textual data requires technical skills to decipher text and generalise concepts and relationships. 
On the other side,  designing a \textit{generic schema} that is not tied to a specific data model -
an abstraction layer that provides a standardised framework for mapping between different database models - requires a thorough understanding of the abstractions involved in \textit{all} database models (e.g., relational, graph).



\paragraph{Main contributions of this paper.}

\begin{itemize}[leftmargin=*]
\item The proposal of an original way for structuring  textual data.
    \item The proposal of an attribute grammar \metaG  as a meta-grammar that sets the global rules that should be respected by a model-agnostic 
  grammar (or schema).  A  grammar $G_T$ respecting \metaG is seen as a model-agnostic database schema which can be then translated to any  data model (e.g., relational, NoSQL).

    \item The formalization of textual data structuring through the evolution process of a grammar $G_0$ (associated to the syntactic trees of the sentences) to a target grammar $G_T$ which respects  the meta-grammar \metaG.

    \item The formalization of an evolution process through transformations on the trees (instances) guided by tree rewriting rules and a similarity measure.
    
    \item A proof-of-concept implementation  that uses clinical cases as input and \metaG as a meta-grammar.
    
   \end{itemize}

\paragraph{Framework.}
Transforming textual data into a database instance requires  altering  abstraction levels  to generalise information when feasible. 
To cope with its complexity, this structuring process should be divided into distinct phases. 
\architxt\ aims to organise textual data into a generic hierarchical structure. 
This structure is considered as an intermediate product, the input of a post-processing phase that can automatically or semi-automatically align it to a specific database model. 
Let us consider a sentence as a small text to illustrate our approach (\emph{SOSY} stands for Sign or Symptom):
\begin{center}
  \small
  $\text{An}\ \overbracket{\text{intravenous}}^{\text{\tiny ANATOMY}}\ \overbracket{\text{urography}}^{\text{\tiny EXAM\_NAME}}\ \text{shows}\ \overbracket{\text{bilateral}}^{\text{\tiny ANATOMY}}\ \overbracket{\text{ureteropyelocal dilatation}}^{\text{\tiny SOSY\_DESC}}$
\end{center}

This sentence is pre-processed to generate a syntax tree  as input to \architxt. 
Then \architxt\ produces a grammar $G_T$ as an output. 
$G_T$  includes the following production rules on non-terminal symbols:
\begin{itemize}
  \item $\metaGroup{Sosy} \to \metaProperty{sosyDesc} \metaProperty{anat}$;
  \item $\metaGroup{Exam} \to \metaProperty{examName} \metaProperty{anat}$ and
  \item $\metaRelation{ExamSosy} \to \metaGroup{Exam} \metaGroup{Sosy}$
\end{itemize}
In a post-processing phase, $G_T$ can be aligned with either :
(i)  a relational database  with  tables 
$EXAM[idExam, examName ,anat]$,\linebreak 
$SOSY[idSosy, sosyDesc,$ $anat]$, and 
$EXAM\_SOSY[idExam, idSosy]$ 
or
(ii) a graph database, where EXAM and SOSY are nodes, and EXAM\_SOSY represents an edge between them.

Thus, \architxt\ proposes to structure text into an intermediate model-agnostic formalism.
Its core concepts are universal across database models, though expressed with different terminology: properties are organised into  groups that relate to each other.
 
\begin{figure}
      \centering
      \includegraphics[width=\linewidth]{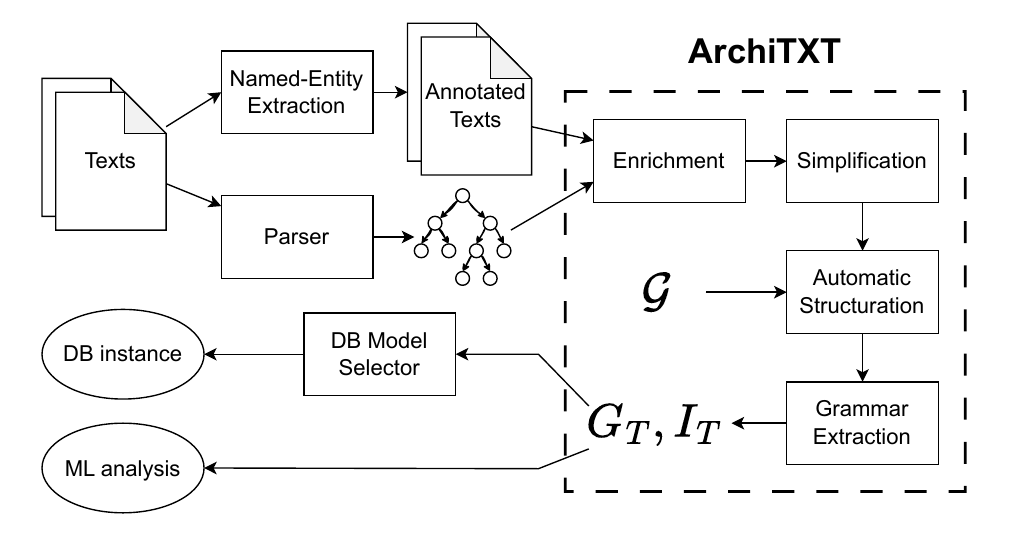}
      \caption{Architecture of \architxt \label{figArchi}}
      \label{architecture}
\end{figure}

Fig.~\ref{figArchi} presents the \architxt\ ecosystem. Input texts are pre-processed using named-entity extraction tools and a parser, preparing them for \architxt.  

\architxt\ generates a structured representation of the data by following a hybrid approach. 
It operates top-down by enriching the tree with named-entities,  extracted during the pre-processing phase by  natural language processing tools.  
It also applies a bottom-up process, analysing the input syntactic tree, pruning or aggregating its sub-trees. 

Thus, in this approach, trees represent database instances, while a grammar capable of generating such trees serves as the database schema. 
More precisely, \architxt\ produces a grammar $G_T$ that defines a generic hierarchical structure representing a model-agnostic database schema, along with its corresponding tree instance, $I_T$, which contains fragments of the original text relevant for storage.
The schema respects  the meta-grammar~\metaG.


These outputs of \architxt\ can be directly analysed by ML tools or used as inputs for other tools that can automatically or semi-automatically align them to  the most suitable database model (relational, graph, etc.).

\paragraph{Rationale.} 
\architxt\ is designed as a textual data structuring tool that promotes transparency, interoperability and trust between users and their applications. Our approach introduces an innovative, auditable 
framework for data modeling. 
As autonomous platforms evolve, \architxt\ maintains its high level of abstraction and flexibility, allowing, for example, \metaG to be enriched with different constraints or concepts. 
This adaptability reinforces its role in building reliable digital ecosystems.

\paragraph{Paper Organization.} 
Sections~\ref{sec:RelW} and~\ref{sec:Back} outline related work  and background concepts, respect.
Section~\ref{sec:autoStruc} presents our structuring method. Section~\ref{sec:MG} defines the attribute grammar.
Before concluding, section~\ref{sec:POC} analysis experiment results.


\section{Related Work}
\label{sec:RelW}
%

This paper presents a methodology for translating data from text-derived syntax trees into a 
model-agnostic framework that can be aligned with different database models. 
Our solution is related to two main research axes: text structuring and data integration.

\noindent
\textit{Text structuring} has mainly been explored within the NLP (Natural Language Processing) domain, often without addressing the challenge of populating a database. This latter task requires a shift in the level of abstraction beyond that required for pure Information Extraction (IE).
This is why our approach involves developing methods to bridge the gap between raw text and structured data, ensuring that the extracted information is both meaningful and actionable within a database context.

In general, setting structure to textual data  can be considered through top-down and bottom-up perspectives \cite{abdullah2023systematic,HFMV24}.
In  top-down approaches, a schema is provided, and the problem is seen as a query on the text to extract or identify \textit{relevant} information.
In \cite{jurafskySpeechLanguageProcessing2008,minardDOINGDEFTCascade2020,valenzuela2020odinson,savaryRelationExtractionClinical2022,cui2024semantic} we find examples of traditional approaches,
while recent trends are shifting towards machine learning (ML) and large language models (LLMs) for extracting named-entities and relationships \cite{li2020survey,keraghel2024survey,dagdelen2024structured}.
They often depend on large annotated corpora for training, which can be both expensive and time-consuming. While ML methods perform well, they may lack the flexibility to handle novel or unexpected data types. 
Bottom-up approaches, known as  \textit{open information extraction} (OpenIE), are  used  in  \textit{ontology learning} to extract terms, named-entities, and relationships from text, classifying and grouping them based on similarity or syntax rules  \cite{liuDBpediaBasedEntityLinking2018,al-aswadiAutomaticOntologyConstruction2020}.
These methods operate without predefined schemas. 
Early techniques employed syntax patterns to extract $\langle \text{subject}, \text{predicate}, \text{object} \rangle$ triples, linking them to knowledge bases~\cite{morinAutomaticAcquisitionSemantic1999,gamalloMappingSyntacticDependencies2002}.
Recent advancements  shifted towards ML techniques \cite{shenProbabilisticModelLinking2014,fanLinkingEntitiesRelations2024} 
to enhance extraction accuracy \cite{sakorFalconEntityRelation2020,liEfficientOnePassEndtoEnd2020,zhou2022survey,zhu2023large,navigliWordSenseDisambiguation2024}.
Despite these developments, ontology learning remains  complex~\cite{browarnikOntologyLearningText2015} and often requires human supervision, although LLMs tend to avoid it.
Current OpenIE models still struggle to extract meaningful relationships and lack a standardized output format~\cite{liu2022open}.


In contrast to our approach, most related work focuses on IE. 
Those related to ontology learning are closer to ours. However, our work focuses on structuring rather than IE, using existing IE tools as sources for our enrichment step.
Our approach is therefore an  hybrid one that combines syntax tree transformation with prior semantic enrichment. 
Hybrid methods have shown potential to increase efficiency: in  Semi-Open IE (SOIE)~\cite{yu2021semi} for domain-independent fact extraction;  in~\cite{wilke2016merging} for merging structured and unstructured information.

\noindent
\textit{Data integration} is often achieved through a common model that maps to different database models, enabling interaction with multiple sources~\cite{barretAbstraGenericAbstractions2022,maliFACTDMFrameworkAutomated2024}.
Some approaches focus solely on providing unified access to data stored in different formats, providing users with a generic, global view of the entire dataset. This view is commonly structured using entity-relationship (E/R) principles. In this context, \cite{GMC21,GMC22} introduce a language for specifying conceptual and physical schemas, along with mapping rules to express their correspondence. 
The E/R idea as a schema over the data set is also used in~ \cite{barretAbstraGenericAbstractions2022} where structured and semi-structured data is viewed through an intermediate structure based on concepts (sub-records, records, collections) similar to those in our generic schema (properties, groups, relations, collections). For this reason, we consider it the work most closely related to ours. 
They map (semi-)structured data from various sources to a graph format, detecting nodes that correspond to the key concepts they define. 
Once nodes have been classified based on some user input, they are given semantic meaning, often using an ontology -- a step that is beyond the scope of our work.
However, unlike our approach, neither \cite{barretAbstraGenericAbstractions2022} nor \cite{GMC21,GMC22} deal with unstructured data or transform data between formats. Instead, they provide a way to explore and navigate datasets to help users understand or access the data.

In~\cite{maliFACTDMFrameworkAutomated2024}, as in our approach, both data and schema are transformed. Similarly, \cite{YL21,YLZY21} propose designing a high-performance relational database from multi-model data and queries using reinforcement learning. Their method starts by shredding data into small tables and iteratively joining  them.
\cite{maliFACTDMFrameworkAutomated2024} presents a framework that derives the most suitable NoSQL model from a conceptual model based on a meta-model and transformation rules. Unlike our approach, which remains model-agnostic, their goal is to tailor a general model to a specific NoSQL or relational model. These ideas could complement our work as a post-processing step: from our target grammar $G_T$ and instance $I_T$, we could determine the most appropriate database model (e.g., a high number of relationships might suggest a graph model).
We are currently finalizing a tool to convert $G_T$ into a relational or graph database.


To take our method a step further and also produce a user-friendly final schema, we might consider how to
\begin{enumerate*}[label=(\arabic*)]
\item \label{i1} assign semantic names to our structures for better clarity (as in~\cite{barretAbstraGenericAbstractions2022})
  and

\item \label{i2} explore (as in~\cite{maliFACTDMFrameworkAutomated2024}) which database model best suits our schema.
Both points are beyond the scope of this paper. 

\end{enumerate*}


\section{Background}
\label{sec:Back}

This section presents concepts and definitions relevant to this work.

 \begin{definition}
\label{def:struct:tree}
An 
\textbf{ordered tree} $T = (D, l)$ 
consists of a domain $D$ and a labelling function $l$. 
The domain $D \subseteq (\mathbb{N})^*$ is a set of integer sequences (e.g., $x.y.z$), and the labelling function $l : D \to \Sigma \cup \{\lambda\}$ maps elements of $D$ to labels in $\Sigma$ or a special symbol $\lambda$. The domain $D$ satisfies:
\begin{enumerate*}[label=(\arabic*)]
    \item $D$ is prefix-closed: if $u' \in D$ and $u$ is a prefix of $u'$, then $u \in D$ ;
    \item If $u.j \in D$, then $u.i \in D$ for all $i \in \mathbb{N}$ with $0 \leq i < j$.
\end{enumerate*}
Each $u \in D$ is a \emph{position}. For a node $n$ at position $p$, its \emph{depth} is $|p|$. The root is at $\epsilon$, labeled $\lambda$, i.e., $l(\epsilon) = \lambda$. An empty tree is $T = (\{\epsilon\}, \langle \epsilon \mapsto \lambda \rangle)$.  The set of all trees is denoted $\mathbb{T}$.
Relations between nodes:
(i) $v \prec u$ if $u = v.i$, where $v$ is the \emph{parent} of $u$, and $u$ is its \emph{child}.
(ii) $v \prec^* u$ if $u = v.v'$, meaning $v$ is a \emph{prefix} of $u$.
(iii) $v \preceq u$ (direct prefix) or $v \preceq^* u$ (indirect prefix) implies $v$ is an \emph{ancestor} of $u$, and $u$ is a \emph{descendant} of $v$.
(iv) A \emph{leaf} is a node $u$ with no children ($u.0 \notin D$). \qed
\end{definition}
\begin{definition}
    Given a tree $T = (D, l)$, a \textbf{sub-tree} of $T$ at position $u \in D$ is denoted by $T|_u = (D', l')$ and has the following properties:
    \begin{enumerate*}
        \item $D' \subseteq D$ such that $\forall v \in D' ~ u \preceq^* v$ and
        \item $l' = \langle v \mapsto l(v) \mid v \in D' \rangle $.
    \end{enumerate*}
    Moreover, if $t = T|_u$ is a sub-tree, we denote by $t' = P_i^t$ the $i$-th 
    tree-ancestor of $t$ when $t' = T|_v$, $u = vw$ and $|w| = i$.
    We note $\mathbb{ST}$ the set of all sub-trees.\qed
\end{definition}


\begin{example}
Fig.~\ref{fig-trees-ex1-stx} illustrates a tree $T = (D, l)$. The positions noted in this figure are the elements of  $D$. 
We have, for instance:
 $l(\epsilon) = \lambda$; $l(0)=NP$ and $l(1.1.1)=NN$.
    Note that  $T|_{1}$,  a sub-tree of $T$, is not a tree because $D'= \{1.1, 1.1.0, 1.1.0.0, 1.1.1, 1.1.1.0\}$ does not respect the conditions of Definition ~\ref{def:struct:tree}. Here, $P_1^{T|_{1}}= T|_1$ and $P_2^{T|_{1}}= T$.
 
\end{example}


A context-free grammar (CFG) is a set of production rules that generate strings in a formal language. It includes non-terminal symbols ($N$) and terminal symbols ($T$). Each rule, $X \to \alpha$, allows a non-terminal $X$ to be replaced by a string $\alpha$ of terminals and non-terminals. Every CFG starts with a designated non-terminal symbol.
\architxt\ generates \textbf{condensed CFG} that allows rules of the form $X \to \alpha^+$, representing repeated occurrences of $\alpha$, equivalent to $X \to \alpha$ and $X \to \alpha~X$. 
A \textbf{parse tree}, a.k.a. 
derivation tree or syntax tree, describes how the starting symbol of a grammar $G$ derives a word in the language.

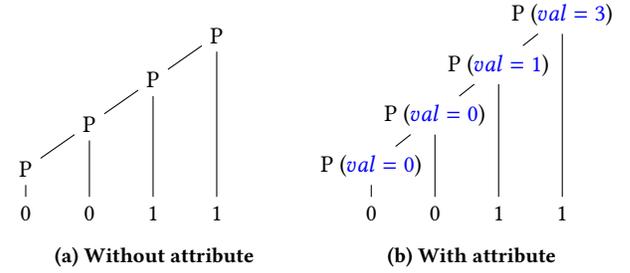
\begin{figure}
    \centering
    \begin{subfigure}{.45\linewidth}
        \begin{adjustbox}{max width=\linewidth,valign=c}
            \begin{forest}
                for tree={calign=last,s sep=1.5em,l sep=.5em,l=1em,fit=tight},
                where n children=0{tier=word}{}
                [P
                    [P
                        [P
                            [P
                                [0]
                            ]
                            [0]
                        ]
                        [1]
                    ]
                    [1]
                ]
            \end{forest}
        \end{adjustbox}
        \caption{Without attribute}
        \label{fig:ex-derivation-1}
    \end{subfigure}
    \hfil
    \begin{subfigure}{.5\linewidth}
        \begin{adjustbox}{max width=\linewidth,valign=c}
            \begin{forest}
                for tree={calign=last,s sep=1.5em,l sep=.5em,l=1em,fit=tight},
                where n children=0{tier=word}{}
                [{P ({\color{blue}$val=3$})}
                    [{P ({\color{blue}$val=1$})}
                        [{P ({\color{blue}$val=0$})}
                            [{P ({\color{blue}$val=0$})}
                                [0]
                            ]
                            [0]
                        ]
                        [1]
                    ]
                    [1]
                ]
            \end{forest}
        \end{adjustbox}
        \caption{With attribute}
        \label{fig:ex-derivation-2}
    \end{subfigure}
    \caption{Example of derivation}
    \label{fig:ex-derivation}
\end{figure}
An attribute grammar extends a CFG by adding semantic information, stored in attributes tied to the grammar's terminal and non-terminal symbols. Attribute values are computed through rules linked to the grammar's productions. For each non-terminal in a CFG $G$, there are two sets of attributes:
\begin{enumerate*}[label=(\arabic*)]
    \item \textit{synthesized attributes}, which pass information from the leaves to the root of a derivation tree and
    \item \textit{inherited attributes}, which pass information from the root to the leaves.
\end{enumerate*}
Each production is paired with semantic rules that compute  synthesized attributes for the left-hand non-terminal and inherited attributes for  the right-hand non-terminals. The formal definition follows.

\begin{definition}
\label{def:struct:G-attr}
An \textbf{attribute grammar}~\cite{knuthSemanticsContextfreeLanguages1968} is a CFG $G = (N, T,$ $R, S)$ augmented with semantic rules $\Phi_r$ for each production $r \in R$.
Each symbol $X \in (N \cup T)$ has attributes $A(X)$ split into:
\begin{enumerate*}[label=(\roman*)]
    \item $A_\uparrow(X)$ for synthesized attributes (empty for terminals $X \in T$), and
    \item $A_\downarrow(X)$ for inherited attributes (empty for $S$).
\end{enumerate*}
Attributes in $A(X)$ have values $V_a$, chosen for each occurrence in the derivation tree.
Production rules $r: X_0 \to X_1, \dots, X_n$ (where $n \ge 1$, $X_0 \in N$, $X_i \in (N \cup T)$) are associated with semantic rules $\varphi \in \Phi_r$,  which are functions defining output $a = \varphi(b_1, \dots, b_k)$ from input attributes $b_i$.
Attributes are synthesized for $X_0$ and inherited for $X_j$ ($1 \leq j \leq n$).
\textbf{S-attributed} grammars contain only synthesized attributes.\qed
\end{definition}

\begin{example}
\label{ex:struct:cfg-attr}
Let $G$ be  a CFG with only one non-terminal $P$ (the start symbol) and terminals $0$ and $1$.
The set of production rules $R$ is
  $\{P  \to 0 \mid 1 \mid  P~0  \mid   P~1\}$.  
  This  grammar produces binary numbers.
  For instance, the parse tree for $0011$ is shown in Fig.~\ref{fig:ex-derivation-1}.
Now, let  $G'$  be an  attribute grammar  built from $G$
by associating  a synthesized attribute $val$ with the non-terminal $P$ and providing rules to compute the value of  $val$ relative to the value of the previously computed attribute $val'$ associated with the right side of the production rule.

\begin{center}
    \small
    \setlength{\abovedisplayskip}{0pt}
    \setlength{\abovedisplayshortskip}{0pt}
    \setlength{\belowdisplayskip}{0pt}
    \setlength{\belowdisplayshortskip}{0pt}
    \begin{minipage}[c]{.3\linewidth}
        \begin{align*}
            P_{val} & \to 0 & [val \gets 0] \\
            P_{val} & \to 1 & [val \gets 1]
        \end{align*}
    \end{minipage}%
    \quad\vline%
    \begin{minipage}[c]{.65\linewidth}
        \begin{align*}
            P_{val} & \to P_{val'}~0 & [val \gets 2 * val'] \\
            P_{val} & \to P_{val'}~1 & [val \gets 2 * val' + 1]
        \end{align*}
    \end{minipage}
\end{center}

In the production rules, the attributes are indicated as subscripts and the semantic rules are presented in square brackets to the right of the production rule.
The grammar $G'$ associates the decimal value with a binary number.
The parse tree for $0011$ in Fig.~\ref{fig:ex-derivation-2} shows (in blue), for each level, the computed value of the attribute $val$.


\end{example}


\begin{definition}
    A \textbf{meta-grammar} $\metaG = (\mathcal{N}, \mathcal{T}, \mathcal{R}, \mathcal{S})$ is an S-attribute grammar where $\mathcal{N}$ is the set of meta-non-terminals, $\mathcal{T}$ the set of meta-terminals, $\mathcal{R}$ the production rules, and $\mathcal{S} \in \mathcal{N}$ the start symbol.
    \metaG defines the syntax of production rules for condensed CFGs, with words in its language representing these rules. A synthesized attribute $\gamma$ ensures derivation validity: $\mathcal{S}_\gamma = \top$ for valid and $\mathcal{S}_\gamma = \bot$ for invalid derivations.
    Semantic rules in $r \in R$ are:
    (i) $a \gets \alpha$, where $a$ is an attribute and $\alpha$ a formula on $r$'s attributes;
    (ii) $\gamma \gets \beta$, where $\beta$ is a logical formula on $r$'s attributes.
    We omit $\gamma \gets$ and trivial rules of the form $a \gets a$.
    \qed
\end{definition}

\begin{example}
    Let's consider a meta-grammar   $\metaG = (\mathcal{N}, \mathcal{T}, \mathcal{R}, \mathcal{S})$
    with  $\mathcal{N} =\{\langle \mathcal{S}_\gamma \rangle\}$, \ie, only one meta-non-terminal (a starting one),
    $\mathcal{T} =\{FirstName, LastName\}$ and the following production rule, where $::=$ is the meta-rule separator. 
    
        \begin{align*}
        \langle \mathcal{S}_{\gamma} \rangle &::= \lambda \to FirstName_{x} ~ LastName_{y} & [\gamma \gets x = y]
    \end{align*}

    A meta-grammar defines a (context-free) grammar. Thus, here, the  left-hand side, $\langle \mathcal{S}_\gamma \rangle$, is a meta-non-terminal, while the right-hand side specifies a grammar.
In this example, the meta-rule defines a grammar with a single production $\lambda \to FirstName ~ LastName$. Moreover, the meta-grammar $\metaG$ is an attribute grammar: the production involves attributes $x$ and $y$, subject to a semantic rule imposed by $\metaG$ ($[\gamma \gets x = y]$).
This semantic rule requires the attributes of $FirstName$ and $LastName$ to match. 
In other words, the  semantic rule ensures that the attributes match correctly (so $FirstName_{stud}$ can only pair with $LastName_{stud}$, not $LastName_{prof}$).
Hence, the production $\lambda \to FirstName_{stud} ~ LastName_{stud}$ is valid (since $stud = stud$), while $\lambda \to FirstName_{stud} ~ LastName_{prof}$ is invalid (since $stud \neq prof$).
$\Box$

\end{example}

Syntax trees (\eg, Fig.~\ref{fig-trees-ex1-stx}) serve as parse trees for the grammar of the natural language. They depict the hierarchical relationships between words or word groups and their roles within parts of speech (PoS) such as nouns (NN), verbs (VBD), adjectives (ADJ), and determiners (DT). The leaves of the tree represent the lexical units (words), while the intermediate nodes correspond to abstract structures like verb phrases (VP) or noun phrases (NP). Our approach applies tree rewriting to transform trees according to specific rules while discarding irrelevant elements.


\section{Automatic Structuring}
\label{sec:autoStruc}
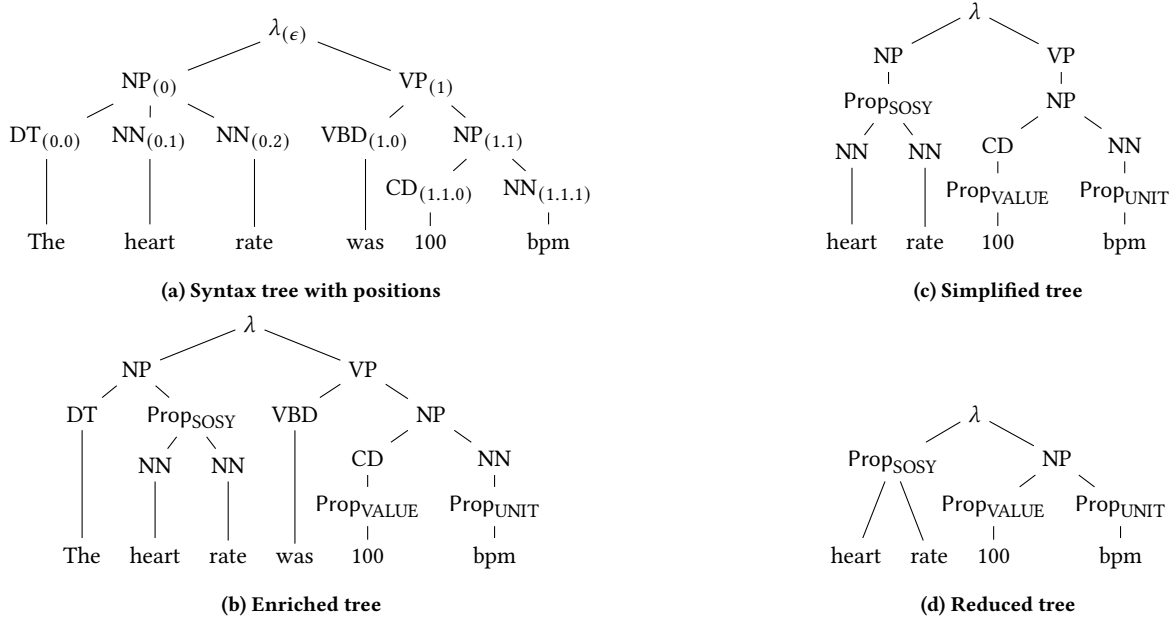
\begin{figure*}[htb]
  \sbox0{
    \begin{subfigure}{.55\linewidth}
      \centering
      \begin{adjustbox}{max width=\linewidth, max height=.2\textheight}
        \begin{forest}
          for tree={s sep=.5em,l sep=.5em,l=1em,fit=tight},
          where n children=0{tier=word}{}
          [$\lambda_{(\epsilon)}$
            [NP$_{(0)}$
              [DT$_{(0.0)}$ [The]]
              [NN$_{(0.1)}$ [heart]]
              [NN$_{(0.2)}$ [rate]]
            ]
            [VP$_{(1)}$
              [VBD$_{(1.0)}$ [was]]
              [NP$_{(1.1)}$
                [CD$_{(1.1.0)}$ [100]]
                [NN$_{(1.1.1)}$ [bpm]]
              ]
            ]
          ]
        \end{forest}
      \end{adjustbox}
      \caption{Syntax tree with positions}
      \label{fig-trees-ex1-stx}
  \end{subfigure}}
  \sbox1{
    \begin{subfigure}{.55\linewidth}
      \centering
      \begin{adjustbox}{max width=\linewidth, max height=.2\textheight}
        \begin{forest}
          for tree={s sep=.5em,l sep=.5em,l=1em,fit=tight},
          where n children=0{tier=word}{}
          [$\lambda$
            [NP
              [DT [The]]
              [\metaProperty{SOSY}
                [NN [heart]]
                [NN [rate]]
              ]
            ]
            [VP
              [VBD [was]]
              [NP
                [CD [\metaProperty{VALUE} [100]]]
                [NN [\metaProperty{UNIT} [bpm]]]
              ]
            ]
          ]
        \end{forest}
      \end{adjustbox}
      \caption{Enriched tree}
      \label{fig-trees-ex1-ent}
  \end{subfigure}}
  \sbox2{
    \begin{subfigure}{.4\linewidth}
      \centering
      \begin{adjustbox}{max width=\linewidth, max height=.2\textheight}
        \begin{forest}
          for tree={s sep=.5em,l sep=.5em,l=1em,fit=tight},
          where n children=0{tier=word}{}
          [$\lambda$
            [NP
              [\metaProperty{SOSY}
                [NN [heart]]
                [NN [rate]]
              ]
            ]
            [VP
              [NP
                [CD [\metaProperty{VALUE} [100]]]
                [NN [\metaProperty{UNIT} [bpm]]]
              ]
            ]
          ]
        \end{forest}
      \end{adjustbox}
      \caption{Simplified tree}
      \label{fig-trees-ex1-simp}
  \end{subfigure}}
  \sbox3{
    \begin{subfigure}{.4\linewidth}
      \centering
      \begin{adjustbox}{max width=\linewidth, max height=.2\textheight}
        \begin{forest}
          for tree={s sep=.5em,l sep=.5em,l=1em,fit=tight},
          where n children=0{tier=word}{}
          [$\lambda$
            [\metaProperty{SOSY}
              [heart]
              [rate]
            ]
            [NP
              [\metaProperty{VALUE} [100]]
              [\metaProperty{UNIT} [bpm]]
            ]
          ]
        \end{forest}
      \end{adjustbox}
      \caption{Reduced tree}
      \label{fig-trees-ex1-reduce}
  \end{subfigure}}
  \centering
  \usebox0 \hfill \usebox2
  \usebox1 \hfill \usebox3
  \caption{Named-entity incorporation in a syntax tree and simplifications (example)}
  \label{fig-trees-ex1}
\end{figure*}

This section explains the structuring process proposed by \architxt.
The approach is grounded from a grammatical perspective: each sentence in the text is associated with its syntax tree, and we consider the initial grammar $G_0$ as the one that accepts these syntax trees.
Fig.~\ref{fig-trees-ex1-stx} shows an example of a syntax tree.

\begin{figure}
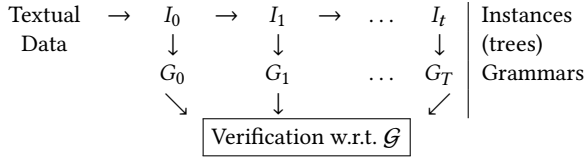

  \centering
  \begin{adjustbox}{max width=.95\linewidth}
    \begin{tabular}{cccccccc|l}
      Textual & $\rightarrow$ & $I_0$                                                 & $\rightarrow$ & $ I_1$       & $\rightarrow$ & $ \dots$ & $I_t$        & Instances \\
      Data    &               & $\downarrow$                                          &               & $\downarrow$ &               &          & $\downarrow$ & (trees)   \\
              &               & $G_0$                                                 &               & $G_1$        &               & $ \dots$ & $G_T$        & Grammars  \\
              &               & ~ $\searrow$                                          &               & $\downarrow$ &               &          & $\swarrow$ ~ &           \\
              &               & \multicolumn{6}{c}{\fbox{Verification w.r.t. \metaG}} &
    \end{tabular}
  \end{adjustbox}
  \caption{Iterative process for automatic structuring}
  \label{fig:struct:general-V0}
\end{figure}

\architxt\ carries out an iterative process that is visually summarized in Fig.~\ref{fig:struct:general-V0}.
We work progressively by transforming the data instance (trees) and the data schema (a grammar).
Our goal is to obtain an instance respecting a target grammar $G_T$, which in turn respects the meta-grammar \metaG.

In Fig.~\ref{fig:struct:general-V0}, the
vertical axis represents the extraction of a grammar from an instance $I_i$ in the form of a rooted forest of enriched syntactic trees, while
the horizontal axis  represents the progression of the process from step $i$ to step $i+1$ based on transformations on the instance.
Successive evolutions  are of two types:
(a)  \textit{Evolution of the instance:} To evolve from instance $I_i$ to instance $I_{i+1}$, the branches of the tree are grouped or transformed based on similarity measures. This may involve reorganising the structure of the sub-trees to improve their consistency or to better align them with the desired representation.
(b)  \textit{Evolution of the grammar:}  The evolution of $G_i$ towards $G_{i+1}$ is triggered by checking whether the grammar $G_i$ conforms to the meta-grammar \metaG.
        At step $i$, if $G_i$ does not conform to \metaG, the process continues by transforming the tree structures of $I_i$, giving rise to the instance $I_{i+1}$, which generates 
        a new grammar $G_{i+1}$.
        The process ends when we find a grammar $G_T$ which satisfies \metaG.


  \begin{algorithm}[htb]
    \caption{\small AlgoStructMain($I_0$ , \metaG, $\textbf{E}$)}
      \label{algo:struct:Main}
      
      $I \gets EnrichSimplify(I_0, E)$ \label{Main-ligne:enrichsimplify} \\
      $G \gets ExtractGrammar(I)$\label{Main-ligne:extract} \\
      
      \While{$G$ not valid wrt \metaG  \label{Main-ligne:while}}{
        $ComputeEqClasses(I)$  \label{Main-ligne:computeeqclasses} \\
                $I \gets Rewrite(I)$ \label{Main-ligne:rewrite}\\
                $G \gets ExtractGrammar(I)$\label{Main-ligne:extract2}
        }   
          
      \Return{$G$}
  \end{algorithm}

Algorithm~\ref{algo:struct:Main} summarises our process for structuring textual data.
As input,  
it receives:
    (1) an instance $I_0$ corresponding to a forest of syntax trees generated from text sources, which have been merged into a single tree with a common root;
    (2) the meta-grammar \metaG and
    (3) a set of named-entities $\textbf{E}$.
A \emph{named-entity} $E$ in natural language processing (NLP) represents a real-world object, denoted by one or more tokens in text, and is an instance of a class (e.g., "Paris" is an instance of "City"). Named-entities are detected in texts and treated as property names in our approach. For example, the pair (City, Paris) is seen as the property "City" with the value "Paris". Algorithm~\ref{algo:struct:Main}  outputs a target grammar $G$ valid with respect to \metaG.




The  process begins with an enrichment step (line~\ref{Main-ligne:enrichsimplify}) where properties (built from the named-entities received as input) are added as internal nodes to the syntax trees, followed by the removal of redundancies to simplify the structure. Next, the grammar is extracted (line~\ref{Main-ligne:extract}) and checked against a pre-established meta-grammar  (line~\ref{Main-ligne:while}). If the resulting grammar is not valid, tree transformations are applied. This involves computing equivalence classes for non-terminals (line~\ref{Main-ligne:computeeqclasses}) and then unifying and structuring equivalent sub-trees according to the meta-grammar \metaG (line~\ref{Main-ligne:rewrite}). 
Then, a new grammar is extracted from the new instance (line~\ref{Main-ligne:extract2}) and the while loop proceeds with verification.

In the rest of this section,  we outline the different steps of Algortihm~\ref{algo:struct:Main}. The definition of the meta-grammar \metaG  is left to Section~\ref{sec:MG}.

\paragraph{\textbf{(A) Transforming Trees.}}

Changes to the instance tree include simplifications, enrichment, and structural transformations. 
The first two occur on line~\ref{Main-ligne:enrichsimplify}, while the latter, guided by \metaG, is performed on line~\ref{Main-ligne:rewrite} of Algorithm~\ref{algo:struct:Main}.

Fig.~\ref{fig-trees-ex1} illustrates an example of transformation on a tree instance.
Fig.~\ref{fig-trees-ex1-stx} shows the syntax tree derived from the sentence \enquote{The heart rate was 100 bpm}. This tree serves as input for \architxt.
The enrichment step consists in incorporating a named-entity (identified in a pre-processing step) in a syntax tree $T$, by creating a \metaProperty{}-labeled sub-tree.
Each sub-tree  specifies the property type (e.g., Person, Country, Disease) and contains leaves for the property's values.
Fig.~\ref{fig-trees-ex1-ent} presents the same tree from Fig.~\ref{fig-trees-ex1-stx}, now enriched with  named-entities, creating nodes \metaProperty{SOSY}, \metaProperty{VALUE}, and \metaProperty{UNIT}. 

Initial simplifications adjust syntax trees to align coordinate conjunctions with English standards. Other steps remove sub-trees without properties (e.g., $T|_{0.0}$, $T|_{1.0}$ in Fig.~\ref{fig-trees-ex1-ent}) and non-property nodes with a single child (e.g., nodes at $0$, $1$, $0.0.0$, $0.0.1$, $1.0.0$, $1.0.1$ in Fig.~\ref{fig-trees-ex1-simp}).
The goal is to eliminate unnecessary terms and nodes, retaining only branches relevant to the database (e.g., the article \enquote{The} in Fig.~\ref{fig-trees-ex1-stx} is removed in Fig.~\ref{fig-trees-ex1-simp}).  
In Fig.~\ref{fig-trees-ex1-reduce}, grammatical nodes 
irrelevant to the database are  removed.

Then, 
each iteration of  Algorithm~\ref{algo:struct:Main} introduces modifications to the tree instance, guided by tree rewriting rules and a similarity measure that helps define equivalence classes.
The rewriting rules aims to prune sub-trees or reorganize them differently.
\textbf{In our current implementation,  we aim to eliminate structural variations, maximize frequency, and minimize the grammar}, but 
the rewriting approach may vary depending on the objectives. 

\begin{figure}
  \centering
  \begin{subfigure}[c]{.5\linewidth}
    \centering
    \begin{adjustbox}{max width=\linewidth, max height=.15\textheight}
      \begin{forest}
        for tree={s sep=.5em,l sep=.5em,l=.5em,fit=tight},
        where n children=0{tier=word}{}
        [$\lambda$
          [\metaCollection{1}
            [\metaRelation{1}
              [\metaGroup{1}
                [\metaProperty{1} [$v_1$]]
                [\metaProperty{2} [$v_2$]]
              ]
              [\metaGroup{2}
                [\metaProperty{3} [$v_3$]]
              ]
            ]
          ]
        ]
      \end{forest}
    \end{adjustbox}
  \end{subfigure}%
  \hfill%
  \begin{subfigure}[c]{.5\linewidth}
    \centering
    \begin{align*}
      \lambda            & \to \metaCollection{1}                  \\
      \metaCollection{1} & \to \metaRelation{1}^+                  \\
      \metaRelation{1}   & \to \metaGroup{1} ~ \metaGroup{2}       \\
      \metaGroup{1}      & \to \metaProperty{1} ~ \metaProperty{2} \\
      \metaGroup{2}      & \to \metaProperty{3}
    \end{align*}
  \end{subfigure}

  \caption{A tree  with internal nodes representing concepts from \metaG and grammar~$G_T$.}
  \label{fig:metaConcepts}
\end{figure}
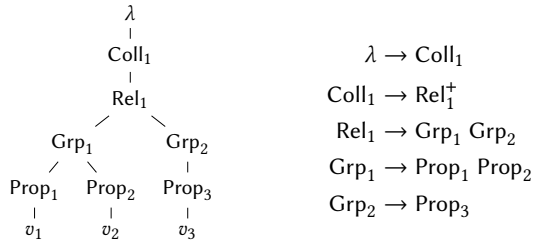

The tree rewriting  function on line~\ref{Main-ligne:rewrite} of Algorithm~\ref{algo:struct:Main}   takes as input an instance $I_i$. 
Called within the while loop (line~\ref{Main-ligne:while}), this function iteratively transforms the instance into a condensed tree that represents the grammar (as we will show in paragraph (D), Fig.~\ref{fig:metaConcepts}).  
Each iteration targets the invalid parts of $I_i$, triggering transformations until the resulting grammar $G_T$ conforms to $\mathbb{G}$ (the termination condition of the while loop).
In practice, to ensure termination, a maximum cycle limit $K$ is set, which may leave some tree parts unresolved.
The whole process is governed by four \textbf{parameters}: $f$ (similarity function), $\tau$ (similarity threshold), $minSup$ (minimum element frequency), and $K$ (maximum cycles).
We refer to~\selfCite{chabinTextDatabasesAttribute2024} for more details on these transformations.

\paragraph{\textbf{(B) Grammar Extraction.}}
In Fig.~\ref{fig:struct:general-V0} (vertical axis), at each step $i$, a grammar $G_i$ is derived from instance $I_i$ through the computation of a \textit{quotient tree}, \ie, a quotient graph with no cycles and no vertices with multiple parents.

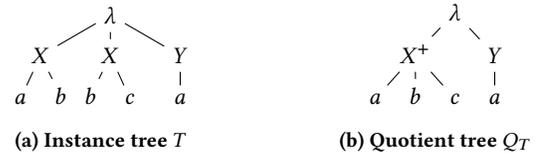
\begin{figure}
    \centering
    \begin{subfigure}{.45\linewidth}
        \centering
        \begin{adjustbox}{max width=\linewidth}
            \begin{forest}
                for tree={s sep=.5em,l sep=.3em,l=1em,fit=tight},
                where n children=0{tier=word}{}
                    [$\lambda$
                        [$X$ [$a$] [$b$]]
                            [$X$,before computing xy={s/.average={s}{siblings}} [$b$] [$c$]]
                            [$Y$ [$a$]]
                    ]
            \end{forest}
        \end{adjustbox}
        \caption{Instance tree $T$}
        \label{fig:struct:quotient:ex:tree}
    \end{subfigure}
    \hfill
    \begin{subfigure}{.5\linewidth}
        \centering
        \begin{adjustbox}{max width=\linewidth}
            \begin{forest}
                for tree={s sep=.5em,l sep=.3em,l=1em,fit=tight},
                where n children=0{tier=word}{}
                    [$\lambda$
                        [$X^+$ [$a$] [$b$] [$c$]]
                            [$Y$ [$a$]]
                    ]
            \end{forest}
        \end{adjustbox}
        \caption{Quotient tree $Q_T$}
        \label{fig:struct:quotient:ex:quotient}
    \end{subfigure}
    \caption{Example of quotient tree}
    \label{fig:struct:quotient:ex}
\end{figure}

The construction of quotient tree can be explained with an example.
First, we retrieve equivalence classes using the equivalence relation $R_l$ on the labels of a tree $T$:
$(\forall x, y \in D) ~ x ~ R_l ~ y \iff l(x) = l(y)$.
We obtain as equivalence classes a set of positions for each label present in the $T$ tree.
In Fig.~\ref{fig:struct:quotient:ex:tree}, the set of equivalence classes $D/R$ includes
 $C_\lambda  = \{\epsilon\}$, $C_X = \{0, 1\}$,  $ C_Y  = \{2\}$,
    $C_a  = \{00, 20\} $, $ C_b  = \{01, 10\}$, $ C_c  = \{11\} $.
Next,  we  \textit{compute the hierarchy of equivalence classes}  which identifies successors of a class $C$ as
$\textsf{Succ}(C) = \{C' \mid \exists u \in C, v \in C' \mbox{ such that } u \prec v \}$, \ie, 
 the set of equivalence classes containing at least one element that is a child of an element in $C$.
Starting with $C_\lambda$, the hierarchy is: 
    $\textsf{Succ}(C_\lambda)  = \{C_X, C_Y\}$,
    $ \textsf{Succ}(C_X)  = \{C_a, C_b, C_c\} $,
    $ \textsf{Succ}(C_Y)  = \{C_a\} $,
    $\textsf{Succ}(C_a)  = \emptyset $,
    $ \textsf{Succ}(C_b)  = \emptyset $,
    $ \textsf{Succ}(C_c)  = \emptyset$.
Finally,  with \textsf{QDom}, we
\textit{assign each equivalence class to  a position in the ordered tree}:
 $\textsf{QDom}(C, u)$ maps each successor of $C$ to a position $u.p$, based on the number of successors $n$: 
    \small $\textsf{QDom}(C, u)=\{(C_j, u.p) \mid C_j \in \textsf{Succ}(C) \mbox{ and } 0 \leq j \leq n-1\}$.
    \normalsize
Starting with $\textsf{QDom}(C_\lambda, \epsilon) = \{(C_X, 0), (C_Y, 1)\}$, successive applications of \textsf{QDom} yield the quotient tree in Fig.~\ref{fig:struct:quotient:ex:quotient}.  
Classes  $C_a$ is duplicated in $Q_T$, because it is linked to two positions: $00$ and $10$ --  it is possible for an equivalence class to appear as the successor of more than one class.
The node for  class $C_X$ in $Q_T$ is marked with `${}^+$', indicating that $X$ can be repeated as a child of $\lambda$ as $C_X = \{0, 1\}$ contains two positions sharing  the parent  $\epsilon$.

A tree $T$ may represent an incomplete derivation of the condensed CFG $G_T$ obtained from a quotient tree $Q_T$.
 If a production rule of $G_T$ is $X \to a~b~c$, the tree $T$ in Fig.~\ref{fig:struct:quotient:ex:tree} is    valid  only if  $c$ or $a$ are treated as missing values. 
 Handling incomplete information remains a challenge in database research~\cite{imielinskiIncompleteInformationRelational1984,Lib06,chabinConsistentUpdatingDatabases2020,chabinManagingLinkedNulls2023}.
 The  transformation of a quotient tree into a grammar is straightforward as illustrated by the following example.
 From $Q_T$ in  Fig.~\ref{fig:struct:quotient:ex:quotient}, we obtain the grammar $G_T$ with the following three rules:  $\lambda  \to X^+ ~ Y ; ~~ X  \to a ~ b ~ c; ~~~Y  \to a.$
 
%


\paragraph{\textbf{(C) Computing Equivalence Classes.}}

\label{eq-class}

Identifying equivalent sub-trees (which correspond to non-terminals of the target grammar) is key
for aggregating information.
Beyond label comparison, sub-tree equivalence depends on context.
We use regular equivalence~\cite{whiteGraphSemigroupHomomorphisms1983}: entities are equivalent if their surrounding relationships are also equivalent
( e.g., two individuals are equivalent (they are  patients) if they are linked to equivalent concepts like a disease or treatment).


\begin{figure}
    \centering
    \begin{adjustbox}{max width=\linewidth}
        \begin{forest}
            for tree={s sep=.5em,l sep=.5em,l=1em,fit=tight}
            [$\lambda$
                [\dots]
                [CONJ,before computing xy={s/.average={s}{siblings}},where n children=0{tier=word}{}
                        [NP$_1$
                            [$X_1$
                                    [\metaProperty{VALUE} [500]]
                                        [\metaProperty{UNIT} [mg]]
                                ]
                                [\metaProperty{DRUG} [Paracetamol]]
                        ]
                        [NP$_2$
                            [$X_2$
                                    [\metaProperty{VALUE} [200]]
                                        [\metaProperty{UNIT} [mg]]
                                ]
                                [\metaProperty{FREQ} [every day]]
                        ]
                ]
                [\dots]
            ]
        \end{forest}
    \end{adjustbox}
    \caption{Extract of an enriched tree}
    \label{fig:struct:sim:ex}
\end{figure}
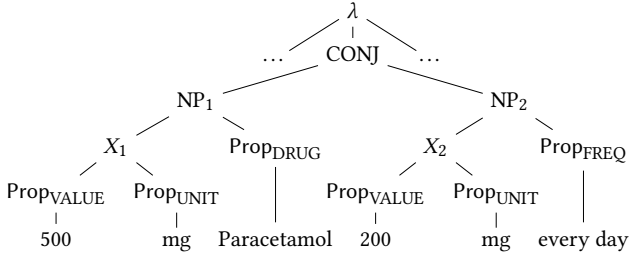

The equivalence relation is defined using a similarity measure  between sub-trees which is a symmetric function $f : \mathbb{ST} \times \mathbb{ST} \to [0,1]$ with $f(x, x) = 1$ for all $x \in \mathbb{ST}$.
Measures  $f$ like Jaccard, Levenshtein, Jaro, or tree edit distance \cite{zhangSimpleFastAlgorithms1989} can be used.
The contextual similarity between two enriched sub-trees $x = T|_u$ and $y = T|_v$, denoted $sim_f(x, y)$, is computed as a weighted average of the recursive similarities provided by the function $f$ for each tree-ancestor.
The weights decrease as the distance from the tree-ancestor increases.
The formula is :
\begin{equation*}
    sim_f(x, y) = \frac{\sum_{i=0}^{depth_{min}} \frac{1}{i + 1} \cdot f(P^x_i, P^y_i)}{\sum_{j=0}^{depth_{min}} \frac{1}{j + 1}}
\end{equation*}
where 
$depth_{min} = \min\{|u|, |v|\}$, 
and $P^x_i$ (or $P^y_i$) is the $i$-th tree-ancestor of $x$ (or $y$).

\begin{example}
    Let $T$ be the tree of Fig.~\ref{fig:struct:sim:ex}.
    To define $f$, we use the Jaccard index ($J(X, Y) = \lvert X \cap Y \rvert / \lvert X \cup Y \rvert$) on the property names present in the sub-tree.
    Although $f(\text{X}_1, \text{X}_2) = 1$, their contexts - one related to a drug (paracetamol) and the other to a frequency (every day) - suggest a similarity of less than $1$.
    Function $sim_f$ 
    accounts for this difference.
    We find $f(\text{NP}_1, \text{NP}_2) = 0.5$ and continue recursively to the root, where the sub-trees are identical with a similarity of $1$.
    The results is :
    \begin{equation*}
       sim_f(\text{X}_1, \text{X}_2) = \frac{\overbrace{1 \times 1}^{\text{X}} +
        \overbrace{0.5 \times 0.5}^{\text{NP}}+
         \overbrace{0.33 \times 1}^{\text{CONJ}} +
          \overbrace{0.25 \times 1}^{\lambda}
        }{1 + 0.5 + 0.33 + 0.25} \simeq 0.88
    \end{equation*}
\end{example}

In the example,  the two sub-trees are similar but, they are not equivalent because they do not refer to the same objects.
Our similarity measure determines the similarity of the sub-trees using a threshold specific to the dataset.

Formally, given sub-trees $x = T|_u$ and $y = T|_v$  of an enriched tree $T$ and a threshold $\tau \in [0, 1]$,
we say that $x$ and $y$ are $\tau$-similar, denoted $x \sim_\tau y$, if and only if $sim_f(x, y) \geq \tau$.
Once we have  a  $\tau$-similarity relation,    we define an equivalence relation between these sub-trees,  
denoted $x \equiv_\tau y$, by the equation :
    \begin{equation*}
      (\forall x, y \in \mathbb{ST}) ~ x \equiv_\tau y \iff x \sim_\tau y \lor (\exists z \in \mathbb{ST}) ~ x \equiv_\tau z \land y \equiv_\tau z
    \end{equation*}

\begin{definition}[Equivalence classes]
    \label{def:eqClasses}
    Let $[x]_\tau$ be  the $\tau$-equivalence class of $x$, where $y \in [x]_\tau$  if and only if $y \equiv_\tau x$.
    For  $T = (D, l)$, $D/_{\equiv_\tau} = \{[x]_\tau \mid x \in D\}$ represents the quotient set (or partition) of $D$ by $\equiv_\tau$, i.e.,  the set of all $\tau$-equivalence classes of~$D$.\qed
\end{definition}

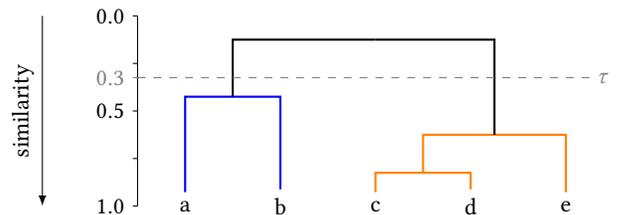
\begin{figure}[htb]
   \centering
   \begin{adjustbox}{max width=\linewidth}
       \begin{tikzpicture}[sloped, x=2em, y=2em]
           \node (a) at (-6,0) {a};
           \node (b) at (-4,0) {b};
           \node (c) at (-2,0) {c};
           \node (d) at (0,0) {d};
           \node (e) at (2,0) {e};
           \node (ab) at (-5,2.3) {};
           \node (cd) at (-1,0.7) {};
           \node (cde) at (.5,1.5) {};
           \node (all) at (-2,3.5) {};

           \draw[thick, blue] (a) |- (ab.center);
           \draw[thick, blue] (b) |- (ab.center);
           \draw[thick, orange] (c) |- (cd.center);
           \draw[thick, orange] (d) |- (cd.center);
           \draw[thick, orange] (e) |- (cde.center);
           \draw[thick, orange] (cd.center) |- (cde.center);
           \draw[thick] (ab.center) |- (all.center);
           \draw[thick] (cde.center) |- (all.center);

           \draw[<-,latex-] (-9,0) -- node[above]{similarity} (-9,4);

           \draw (-7,0) -- (-7,4);

           \foreach \y in {0,1,2,3,4}
           \draw[shift={(0,\y)},color=black] (-7,0) -- (-7.1,0);

           \node[left] at (-7.1,0) {$1.0$} ;
           \node[left] at (-7.1,2) {$0.5$} ;
           \node[left] at (-7.1,4) {$0.0$} ;

           \draw[dashed,color=gray] (-7,2.7) -- node[at end, right]{$\tau$} (2.5,2.7);
           \draw[color=gray] (-7,2.7) -- (-7.1,2.7) node[left] {$0.3$};
       \end{tikzpicture}
   \end{adjustbox}
   \caption{Example of a dendrogram}
   \label{fig:dendrogram}
\end{figure}

\begin{remark}
    According to \cite{carlssonCharacterizationStabilityConvergence2010}, equivalence-based partitioning with a similarity function is equivalent to single-link hierarchical clustering. 
    The process begins with each element in its own class, progressively merging the closest pairs of class based on the maximum pairwise similarity, and stops when the maximum pairwise similarity falls below $\tau$.
    Figure~\ref{fig:dendrogram} illustrate the idea.
\end{remark}

\paragraph{\textbf{(D) Model-agnostic Schema.}}

If our goal is model-agnostic database design, we must identify general abstractions that are common to all database models.
\architxt\ structures data using four generic key concepts that can be aligned with different database models:
(1) \emph{Property} (\metaProperty{}). A named data value defining an entity's characteristics, called attributes in relational models, properties in graphs, and fields in documents.
(2) \emph{Group} (\metaGroup{}). A named set of attributes representing the main data unit, called tuples in relational models, nodes in graphs, and documents in document stores.
(3) \emph{Relation} (\metaRelation{}). A relationship between distinct groups,  representing entity relationships - foreign keys in relational models, edges in graphs, or embedded references in documents.
(4) \emph{Collection} (\metaCollection{}). A set of equivalent groups and relations, abstracting data organization as set of tables with foreign key relationships (relational), subgraphs (graph), or document collections sharing a common structure or references (document stores).

\noindent

Our model-independent schema is built according to these concepts and formalised by the meta-grammar \metaG.
In \architxt, tree generalizations and transformations are guided by \metaG, aiming to derive an instance tree $I_T$ that adheres to a target grammar $G_T$, which itself conforms to \metaG.
Fig.~\ref{fig:metaConcepts} illustrates the outputs of \architxt.


\section{Meta-Grammar: generic database schema definition}
\label{sec:MG}
\begin{figure}
  \centering
  \begin{adjustbox}{max width=\linewidth}
    \begin{forest}
      for tree={s sep=.3em,l sep=.5em,l=1em,fit=tight,align=center},
      [$\epsilon$
        [{$\langle root \rangle$\\$\color{blue}crL=\{1\}$}
          [$\lambda$]
          [$\to$,before computing xy={s/.average={s}{siblings}}]
          [{$\langle rootList \rangle$\\$\color{blue}crL'=\{1\}$}
            [\metaCollection{1}]
            [$\langle rootList \rangle$
              [$\epsilon$]
            ]
          ]
        ]
        [\textsc{eol},before computing xy={s/.average={s}{siblings}}]
        [{$\langle ruleList \rangle$\\$\color{blue}crL=\{1\}$, $\color{blue}gL=\{1, 2\}$\\$\color{blue}rL=\{1\}$, $\color{blue}pL=\{1, 2, 3\}$}
          [{$\langle collRel \rangle$\\$\color{blue}x=1,r=1$}
            [\metaCollection{1}]
            [$\to$,before computing xy={s/.average={s}{siblings}}]
            [$\metaRelation{1}^+$]
          ]
          [\textsc{eol},before computing xy={s/.average={s}{siblings}}]
          [\dots]
        ]
      ]
    \end{forest}
  \end{adjustbox}
  \caption{Extract of a derivation of \metaG}
  \label{fig:metaDerivation}
\end{figure}
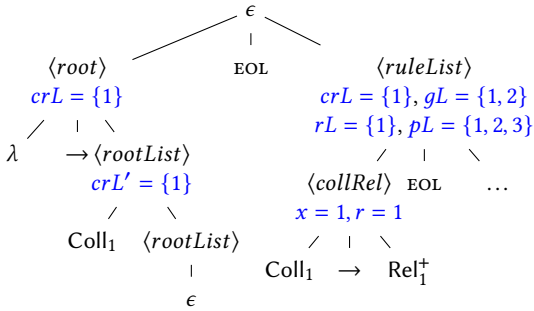

The purpose of the meta-grammar is to define the core concepts of our model-agnostic database schema. These concepts -- property, group, relation, and collection -- must now be expressed   by trees to enable their identification within the tree structure representing our data instance $I$.
Table~\ref{table:struct} presents our meta-grammar \metaG, an attribute grammar (Definition~\ref{def:struct:G-attr}) that defines valid data structures. Meta-non-terminals are indicated by angle brackets $\langle \cdot \rangle$, while the semantic rules are shown on the right side of the table within square brackets $[ \cdot ]$.

The first production meta-rule of \metaG (\ref{meta:start}) indicates that the target grammar $G$ is defined by an initial rule,  followed by a possibly empty list of rules.
The initial rule generated by \metaG (meta-rule~\ref{meta:root}) contains the symbol $\lambda$ (initial non-terminal of $G$) on its left-hand side.
Its right-hand side is defined by the meta-rules \ref{meta:rootList:1}-\ref{meta:rootList:6} which specify the construction of a series of $G$ non-terminals.
These non-terminals are: \metaProperty{}, \metaGroup{}, \metaRelation{} and \metaCollection{}, representing, respectively, properties, groups of properties, relations between groups and collections of groups or relations.
To distinguish each structure specific to a grammar $G$, we associate a $name$ attribute with each non-terminal.
The attributes of \metaG are synthesized and represent lists of names ($name$) used as identifiers: $pL$ (or $pL'$) denotes  the lists of property names, $gL$ (or $gL'$) the lists of group names, $rL$ (or $rL'$) the lists of relation names, $cgL$ (or $cgL'$) the lists of group collection names, and $crL$ (or $crL'$) the lists of relation collection names.
They are initialized in a bottom-up fashion.
Syntax rules are used to check that a name is unique.
For example, if the meta rules \ref{meta:entList:1}-\ref{meta:entList:2} are applied, the list $pL$ gets property names that are unique.
This is also the case for all other lists: the uniqueness of the name of a new non-terminal is guaranteed by the semantic rules.
It is also important to note that these semantic rules ensure that any non-terminal appearing to the right of a production rule in the grammar $G$  must have a rule defining it.
For example, if the meta rule \ref{meta:start} is applied, all elements in the list $gL'$ must be present in $gL$.\\

\begin{table*}[htb]
  \centering
  \begin{adjustbox}{max width=\linewidth,valign=c}
    \setlength{\abovedisplayskip}{0pt}
    \setlength{\abovedisplayshortskip}{0pt}
    \setlength{\belowdisplayskip}{0pt}
    \setlength{\belowdisplayshortskip}{0pt}
    \begin{minipage}{\textwidth}
      \centering
      \hrule%
      \begin{align}
        & \epsilon ::= \langle root_{pL',gL',rL',cL'} \rangle ~\textsc{eol}~ \langle rules_{pL,gL,rL,cL} \rangle  & [pL' \subseteq pL; gL' \subseteq gL; rL' \subseteq rL; cL' \subseteq cL]         \label{meta:start}      \\
        & \langle root_{pL,gL,rL,cL} \rangle ::= \lambda \to \langle rootList_{pL,gL,rL,cL} \rangle                                                                                                                                                           \label{meta:root}       \\
        & \langle rootList_{pL,gL,rL,cL} \rangle ::= \epsilon                                                                                                 & [pL \gets \emptyset; gL \gets \emptyset; rL \gets \emptyset; cL \gets \emptyset] \label{meta:rootList:1} \\
        & ~~ \mid ~ \metaProperty{x}  ~ \langle rootList_{pL',gL,rL,cL} \rangle                                         & [x \notin pL'; pL \gets \{x\} \cup pL']                                                          \label{meta:rootList:2} \\
        & ~~ \mid ~ \metaGroup{x} ~ \langle rootList_{pL,gL',rL,cL} \rangle                                        & [x \notin gL'; gL \gets \{x\} \cup gL']                                                          \label{meta:rootList:3} \\
        & ~~ \mid ~ \metaRelation{x}  ~ \langle rootList_{pL,gL,rL',rL} \rangle                                         & [x \notin rL'; rL \gets \{x\} \cup rL']                                                          \label{meta:rootList:4} \\
        & ~~ \mid ~ \metaCollection{x} ~ \langle rootList_{pL,gL,rL,cL'} \rangle                                         & [x \notin cL'; cL \gets \{x\} \cup cL']                                                       \label{meta:rootList:6} \\
        & \langle rules_{pL,gL,rL,cL} \rangle ::= \epsilon                                                                                                 & [pL \gets \emptyset; gL \gets \emptyset; rL \gets \emptyset; cL \gets \emptyset] \label{meta:ruleList:1} \\
        & ~~ \mid ~ \langle property_{x}          \rangle ~\textsc{eol}~ \langle rules_{pL',gL,rL,cL} \rangle & [x \notin pL'; pL \gets \{x\} \cup pL']                                                          \label{meta:ruleList:2} \\
        & ~~ \mid ~ \langle group_{x, pL'}      \rangle ~\textsc{eol}~ \langle rules_{pL,gL',rL,cL} \rangle & [x \notin gL' \land pL' \subseteq pL; gL \gets \{x\} \cup gL']                                   \label{meta:ruleList:3} \\
        & ~~ \mid ~ \langle relation_{x, gL'}   \rangle ~\textsc{eol}~ \langle rules_{pL,gL,rL',cL} \rangle & [x \notin rL' \land gL' \subseteq gL; rL \gets \{x\} \cup rL']                                   \label{meta:ruleList:4} \\
        & ~~ \mid ~ \langle collGrp_{x,g} \rangle ~\textsc{eol}~ \langle rules_{pL,gL,rL,cL'} \rangle & [x \notin cL' \land g \in gL; cgL \gets \{x\} \cup cL']                                  \label{meta:ruleList:5} \\
        & ~~ \mid ~ \langle collRel_{x,r} \rangle ~\textsc{eol}~ \langle rules_{pL,gL,rL,cL'} \rangle & [x \notin cL' \land r \in rL; crL \gets \{x\} \cup cL']                                  \label{meta:ruleList:6} \\
        & \langle group_{x, pL} \rangle            ::= \metaGroup{x} \to \langle propList_{pL} \rangle                                                                                                                                                                     \label{meta:group}      \\
        & \langle collGrp_{x, g} \rangle      ::= \metaCollection{x} \to \metaGroup{g}^+                                                                                                                                                                                 \label{meta:collGroup}  \\
        & \langle relation_{x, gL} \rangle         ::= \metaRelation{x} \to \metaGroup{g1} ~ \metaGroup{g2}                                                             & [g1 \neq g2; gL \gets \{g1, g2\}]                                                          \label{meta:rel}        \\
        & \langle collRel_{x, r} \rangle      ::= \metaCollection{x} \to \metaRelation{r}^+                                                                                                                                                                                   \label{meta:collRel}    \\
        & \langle propList_{pL} \rangle                ::= \metaProperty{x}                                                                                               & [pL \gets \{x\}]                                                                                    \label{meta:entList:1}  \\
        & ~~ \mid ~ \metaProperty{x} ~ \langle propList_{pL'} \rangle                                                         & [x \notin pL'; pL \gets \{x\} \cup pL']                                                          \label{meta:entList:2}  \\
        & \langle property_{x} \rangle               ::= \metaProperty{x} \to \langle data \rangle \label{meta:entity}
      \end{align}%
      \hrule
    \end{minipage}
  \end{adjustbox}
  \caption{Meta-grammar \metaG using BNF format}
  \label{table:struct}
\end{table*}

\begin{example}
  \label{ex:struct}
  Consider the grammar $G$ from Fig.~\ref{fig:metaConcepts} and the derivation of \metaG that leads to the rule $\lambda \to \metaCollection{1}$.
  By applying meta-rule~\ref{meta:root}, we derive the rule $\lambda \rightarrow \langle rootList \rangle$, where the right-hand side contains a meta-non-terminal.
  Next, applying meta-rule~\ref{meta:rootList:6} results in the intermediate rule $\lambda \rightarrow \metaCollection{1} ~ \langle rootList \rangle$, and finally, meta-rule~\ref{meta:rootList:1} produce $\lambda \rightarrow \metaCollection{1}$.
  The set of production rules for the grammar $G$ is defined by meta-rules~\ref{meta:ruleList:1}-\ref{meta:ruleList:6}, where each rule introduces a non-terminal for $G$.

  Fig.~\ref{fig:metaDerivation} presents a partial derivation of $G$ from the meta-grammar \metaG.
  Attributes are displayed in blue, excluding $\gamma$ and empty-set attributes for clarity.
  Note that the semantic rules of \metaG impose constraints to ensure  that every non-terminal in the target grammar $G$ is properly defined, a requirement for constructing a valid grammar.
  In this context, Fig.~\ref{fig:metaDerivation} shows, on the left-hand side of the root, that  $crL' = \{1\}$, signifying that \metaCollection{1} is referenced in the root rule. On the right-hand side, $\langle ruleList \rangle$ holds $crL = \{1\}$.
  Since $crL' \subseteq crL$, this confirms that every non-terminal appearing in the root rule of $G$ has a corresponding defined production rule, ensuring a valid derivation.
  
  
\end{example}

\section{Proof of concept}
\label{sec:POC}
Our approach is not directly comparable to existing literature.
We evaluate our prototype in two steps. 
First, by  assessing its alignment with pre-defined objectives and second by
analyzing its performance using metrics we will introduce.

We tested our structuring method on a proof-of-concept use case using the CAS corpus~\cite{grabarCASFrenchCorpus2018}, which contains clinical cases. We selected $100$ documents ($1800$ sentences) with $8098$ manually annotated named entities across $10$ categories, some of which are nested.
The initial grammar  has 
$3383$ production rules.

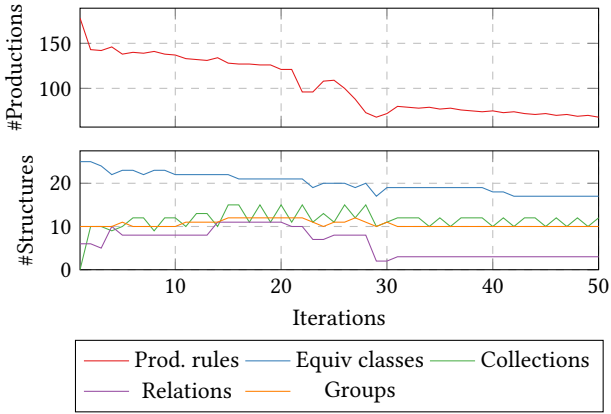
\begin{figure}[htb]
  \centering
  \begin{tikzpicture}
    \begin{groupplot}[
        group style={
          group size=1 by 2,
          xlabels at=edge bottom,
          xticklabels at=edge bottom,
          vertical sep=1em
        },
        xmin=1,
        xmax=50,
        xlabel={Iterations},
        width=\linewidth,
        height=10em,
        grid=major,
        grid style=dashed,
        legend style={
            at={(0.5,-1.2)},
            anchor=south,
            legend columns=3,
        }
    ]
      \nextgroupplot[ylabel={\#Productions}]

      \addplot+[cycle list shift=0] table [x expr={1 + \thisrow{step}}, y=value, col sep=comma] {nb_prod.csv}; \label{figure:struct:xp:prod:prod}

      \nextgroupplot[ylabel={\#Structures}, ymin=0]

      \addplot+[cycle list shift=0, draw=none] coordinates {(0,0)};
      \addlegendentry{Prod. rules};

      \addplot+[cycle list shift=1] table [x expr={1 + \thisrow{step}}, y=value, col sep=comma] {nb_equiv_subtrees.csv}; \label{figure:struct:xp:ratio:equiv}
      \addlegendentry{Equiv classes};

      \addplot+[cycle list shift=2] table [x expr={1 + \thisrow{step}}, y=value, col sep=comma] {nb_coll.csv}; \label{figure:struct:xp:nbElems:coll}
      \addlegendentry{Collections};

      \addplot+[cycle list shift=3] table [x expr={1 + \thisrow{step}}, y=value, col sep=comma] {nb_rel.csv}; \label{figure:struct:xp:nbElems:rel}
      \addlegendentry{Relations};

      \addplot+[cycle list shift=4] table [x expr={1 + \thisrow{step}}, y=value, col sep=comma] {nb_group.csv}; \label{figure:struct:xp:nbElems:group}
      \addlegendentry{Groups};
    \end{groupplot}
  \end{tikzpicture}
  \caption{Evolution of the grammar size during a rewrite of $100$ sentences.}
  \label{figure:struct:xp}
\end{figure}

In \textbf{the first step of our evaluation}, we analyse:
\begin{enumerate*}[label=(\arabic*)]
  \item \textit{Rule reduction}: $G_T$ should be more concise than $G_0$, reflecting aggregation.\label{item:1}
  \item \textit{Hierarchy evolution}: we monitor higher-level generalizations, \eg, how \emph{collections} aggregate \emph{groups}. \label{item:2}
  \item \textit{Core structure instances}: we track the growth of \emph{group}, \emph{relation}, and \emph{collection} instances across iterations. \label{item:3}
\end{enumerate*}
We evaluate these aspects on a reduced example of our corpus with $100$ sentences.
For \ref{item:1}, Algorithm~\ref{algo:struct:Main}  (Fig.~\ref{figure:struct:xp} (top)) reduces production rules from $178$ to $68$. It alternates between specialization and generalization, with steps $24$, $30$, and $31$ introducing specific structures before broader aggregation.
For \ref{item:2},  in Fig.~\ref{figure:struct:xp}(bottom), initially, instance $I_0$ contains $10$ groups, $6$ relations, and no collections. 
By iteration $28$, it reaches $11$ groups, $8$ relations, and $15$ collections, minimizing production rules. 
Subsequent merging reduces relations to $2$, with a third added at iteration $31$ as groups decrease from $15$ to $10$. The process concludes with $10$ groups, $3$ relations, and $12$ collections, stabilizing after iteration $31$, indicating a structured representation of the data. We also notice  that equivalence classes decrease from $25$ to $17$, indicating structural unification. 
Finally, for \ref{item:3}, the average number of group and relation instances remains relatively stable. At the beginning, the average is 34.2 instances per group and 2.2 instances per relation. At the end it is 29.8 instances per group and 3 instances per relation. Indeed, as the number of groups increases, the number of instances per group decreases, indicating that splitting is occurring.

The results of this first analysis are  promising: the obtained generic schema is coherent and reliable.
The target grammar $G_T$, obtained from  the CAS corpus, is valid with respect to \metaG and consists of 12 collections, 3 relations, 10 groups and 9 properties.
Algorithm~\ref{algo:struct:Main}  effectively identifies frequent elements but is highly sensitive to parameters like $minSup$ and $\tau$.
It is important to understand that our method relies primarily on the frequency of syntactic structures. 
Only the enrichment step provides semantic information.
As a result, the scheme remains coherent, but there are differences with human analysis.\\



In \textbf{the second evaluation step}, we assess the effectiveness of \architxt\ by comparing it to a naive baseline approach where each sentence forms a distinct group, with duplicate properties removed. Two groups are considered similar if they share the same property names. For example,  the sentence mentioned in Section~\ref{sec:Intro} implies 1 group with 3 properties: anatomy, exam, and sosy.

Since $G_T$ lacks a gold standard for direct comparison, we define a set of metrics to evaluate schemas.
These metrics cover three distinct aspects:

\noindent
\textbf{Data Loss}, evaluated by a \emph{coverage score} (\cov), measures the proportion of retained property trees, with effective structuring maximising preservation.


\noindent
\textbf{Semantic Loss} is measured by \emph{cluster-adjusted mutual information} (\AMI), which assesses the similarity between two clusterings of a dataset, and \emph{cluster completeness} (\comp), which ensures all data points of a given class are in the same cluster. These metrics evaluate the stability of equivalence classes between original and rewritten trees without penalizing merging. We compare initial clusters derived from parent equivalence classes (Section~\ref{sec:autoStruc}(A)) with final clusters using parent labels post-structuring. Properties absent from final trees form separate clusters. To preserve similarity, these metrics should be maximized.


\noindent
\textbf{Schema Quality} is assessed through three metrics: (1) the \emph{number of production rules} in $G_T$ (\nbR), aiming for schema minimization; (2) \emph{group overlap} (\gO), measured via the Jaccard index on property-name sets, which should be minimized for better information separation; and (3) \emph{redundancy}, where a lower score indicates better normalization, and a higher score suggests the need for further decomposition. Redundancy reflects unnecessary data duplication, with tuples considered redundant if normalization can eliminate repeated information.

Next, we introduce the necessary notions for defining the \textit{redundancy score}  ($\redud$). The idea is to identify functional dependencies within a group, suggesting that data can be further decomposed into smaller, structured groups. This process is analogous to itemset rule mining, where dependencies are evaluated using confidence scores. Functional dependencies are determined at the property level, with their confidence score computed as the average confidence of individual value sets.

We draw a parallel with the relational model, viewing group collections  as tables, to introduce the following definitions: $R[U]$ represents a relation (or table) schema $R$ with a set of attributes $U$. We consider $X$ and $Y$ as disjoint subsets of $U$.

\begin{definition}[Support and Confidence Score]
For a candidate dependency $X \to Y$ and tuple values $x, y$, the \emph{support} of $x$ and the \emph{joint support} of $(x,y)$ are:
$\supp(x) = \left| \{ t \in R \mid t[X] = x \} \right|$ and $\supp(x,y) = \left| \{ t \in R \mid t[X] = x   \text{ and }   t[Y] = y \} \right|$.
 The \emph{confidence} of the instantiated rule $x \to y$ is then given by $\conf(x \to y) = \supp(x,y) / \supp(x)$.
  The \emph{confidence score} of the dependency $X \to Y$ in $R$ is defined as $\conf(X \to Y) = 
  \operatorname{median} \{\conf(x \to y) \mid (x,y) \in \pi_{X \cup Y}(R)\}$.
\end{definition}

%


\begin{definition}[Dependency Score]
For $|X| \geq 2$, the \emph{dependency score} $\depScore(X) = \max \left\{\text{conf}((X \setminus \{A\}) \to A) \mid  A \in X\right\}$ quantifies the strongest functional dependency within $X$.
\end{definition}

\begin{definition}[Redundancy Score]
Let $\alpha \in [0,1]$ be a dependency threshold.
The set of attributes $X$ exhibits \emph{redundancy} in $R[U]$ if $\depScore(X) \geq \alpha$ and $ |\pi_X(R)| < |R|$.
Define $\mathcal{C} = \{ X \subseteq U \mid |X| \geq 2 \text{ and } \depScore(X) \geq \alpha \}$ as the collection of dependent attribute sets.
The \emph{redundancy score} $\redud_\alpha(R)$ measures the fraction of tuples redundant in at least one attribute set $X \in \mathcal{C}$, given by: $\redud_\alpha(R) = \left| \bigcup_{X \in \mathcal{C}} \dup(X) \right| / \left|R\right|$ where $\dup(X) = \{ t \in R \mid \exists t' \in R, t \neq t' \text{ and } t[X] = t'[X] \}$ is the set of redundant tuples with respect to $X$.
\end{definition}

\begin{example}
Let $(a,b,c)$, $(a,b,c')$, $(a',b,c')$ and $(a'',b',c)$ be tuples on $R[ABC]$.
Given $\supp(a) = 2$, $\supp(b) = 3$, and $\supp(a,b) = 2$, we get $\conf(A \rightarrow B) = 1$ and $\conf(B \rightarrow C) = 0.5$.
Thus, $\depScore(AB) = 1$ and $\depScore(BC) = 0.5$.
For $\alpha = 1.0$, $AB$ has redundancy, unlike $BC$.
Finally, we have $\redud_{1.0}(R) = 0.5$

\end{example}

\begin{table}[htb]
  \centering
  \begin{adjustbox}{max width=\linewidth}
    \setlength{\tabcolsep}{.5em}
    \begin{filecontents*}{results.csv}
Run name,Corpus,Sample,Metric,Tau,Support,Coverage,Similarity,Edit distance,Redundancy(1.0),Redundancy(0.7),Redundancy(0.5),CMI,Completeness,Productions,Overlap,Balance
Naive,CAS,full,,nan,nan,0.735,0.476,13540,0.193,0.193,0.193,0.223,0.548,128,0.295,0.646
\architxt,CAS,full,jaro,0.5,10,0.932,0.256,23280,0.463,0.463,0.463,0.138,0.483,14,0.202,0.75
\architxt,CAS,full,jaro,0.5,20,0.853,0.254,22682,0.269,0.308,0.308,0.16,0.509,14,0.262,0.796
\architxt,CAS,full,jaro,0.7,10,1,0.259,23747,0.272,0.292,0.292,0.113,0.425,15,0.194,0.295
\architxt,CAS,full,jaro,0.7,20,0.933,0.255,23340,0.293,0.293,0.293,0.121,0.483,8,0.245,0.386
equiv class,CAS,full,jaro,0.7,10,0.932,0.256,23280,0.463,0.463,0.463,0.144,0.491,14,0.202,0.75
sub-groups,CAS,full,jaro,0.7,10,1,0.259,23954,0.45,0.45,0.45,0.166,0.456,14,0.173,0.59
merge groups,CAS,full,jaro,0.7,10,0.932,0.256,23280,0.463,0.463,0.463,0.144,0.494,14,0.202,0.75
sub-groups + merge,CAS,full,jaro,0.7,10,1,0.259,24069,0.287,0.306,0.306,0.185,0.453,16,0.178,0.378
\end{filecontents*}
\pgfplotstabletypeset[
      fixed,
      zerofill,
      precision=2,
      col sep=comma,
      columns={Run name,Tau,Support,Coverage,Redundancy(1.0),Redundancy(0.5),CMI,Completeness,Productions,Overlap},
      columns/Run name/.style={string type, column name=Algorithm, column type={|c|}},
      columns/Tau/.style={column name=$\tau$, clear infinite, precision=1},
      columns/Support/.style={column name=$minSup$, clear infinite, int detect},
      columns/Coverage/.style={column name=$\cov$},
      columns/CMI/.style={column name=$\AMI$},
      columns/Completeness/.style={column name=$\comp$},
      columns/Productions/.style={column name=$\nbR$},
      columns/Redundancy(1.0)/.style={column name=$\redud_{1.0}$},
      columns/Redundancy(0.5)/.style={column name=$\redud_{0.5}$},
      columns/Overlap/.style={column name=$\gO$},
      every head row/.style={
        before row={\hline & \multicolumn{2}{c|}{Parameters} & \multicolumn{7}{c|}{Results}\\},
        after row=\hline,
      },
      every last row/.style={after row=\hline},
      every column/.append style={column type={c|}},
      skip rows between index={5}{1000},
    ]{results.csv}
  \end{adjustbox}
  \caption{Comparative results}
  \label{table:results}
\end{table}


Table~\ref{table:results} compares the naive approach with \architxt\ across varying values for $\tau$ (similarity threshold) and $minSup$ (minimum element frequency) parameters (Section~\ref{sec:autoStruc}(A)).

Our results show that \architxt\ achieves a very good quality structure for textual data. Notably, the results in columns \nbR\ and \cov\ show that our method significantly reduces the number of structures needed and improves coverage, representing nearly all named-entities annotated in the text.
The production rules in \nbR\ represent non-terminals, each corresponding to groups or relations.
While the naive approach scores slightly higher in \AMI\ and \comp, our method remains competitive.
The difference stems from our more aggressive rewriting process, particularly in identifying subgroups, which can split existing classes into subclasses, affecting the overall score.
The naive approach is not able to achieve a perfect score ($1$) because it misses named-entities that form separate clusters.
Both methods exhibit similar overlap, with ours resulting in a slightly lower value.
\architxt\ achieves a low redundancy score, indicating it effectively identifies meaningful groups despite sentence diversity even if it does not use any semantic information. 




In summary, \architxt\ provides a reliable approach to structuring textual data. By using syntax as its foundation, it effectively identifies meaningful groups and maintains competitive performance metrics. Future enhancements, such as the incorporation of normalisation, could further refine its capabilities.

\paragraph{Reproducibility.}
The procedures are detailed in~\selfCite{hiotConstructionAutomatiqueBases2024}.
The Python implementation uses Stanford CoreNLP for parsing.
The source code for reproducing the experiments is on GitHub~\githubCite{}.
Data access modalities for the corpus are in~\cite{grabarCASFrenchCorpus2018}.

\section{Concluding Remarks}
\label{sec:conclusion}
\architxt\ introduces a novel method for automatically structuring textual data into a flexible, model-agnostic format that abstracts over various database models.
Guided by a meta-grammar \metaG\ - a formal and technically sound tool - our approach bridges unstructured language and structured representations through grammar evolution, driven by similarity measures and tree transformation, without relying on training data. This hybrid, formal strategy ensures  \textit{robustness}, \textit{transparency}, and \textit{auditability}.
Our contribution is both \textit{original} and \textit{flexible}, promoting interoperability across data models and offering a foundation \textit{generalizable} to various domains, possibly with domain-specific adaptations.   
Ongoing work includes incremental structuring and enriching \architxt\ with functional dependency discovery, following approaches such as \cite{papenbrockFunctionalDependencyDiscovery2015}.

\begin{acks}
This work was partially supported by the \textit{Ambition Recherche Développement Centre-Val de Loire} (ARD) JUNON-DATA project.\\
The authors are deeply grateful to the anonymous reviewers for their  constructive feedback and valuable suggestions.
\end{acks}
%
%
%
\bibliographystyle{ACM-Reference-Format}
\bibliography{biblio}
\end{document}